\documentstyle[preprint,aps]{revtex}

\def\CA{{\cal A}}
\def\CO{{\cal O}}
\def\rp{{\not\!\! R_p}}
\def\bd{{B_d^0}}
\def\bdb{{\bar B_d^0}}
\def\GeV{{\:\rm GeV}}
\def\mn{{m_{\tilde \nu_i}}}
\def\me{{m_{\tilde e_i}}}
\def\mb{{m_{\tilde b_R}}}
\def\ms{{m_{\tilde s_R}}}
\def\md{{m_{\tilde d_R}}}
\def\vcd{{V_{cd}}}

\def\vcb{{V_{cb}}}
\def\vtd{{V_{td}}}

\def\vtb{{V_{tb}}}

\def\be{\begin{equation}}
\def\ee{\end{equation}}
\begin{document}

\preprint{UW/PT 97-5}
\title{Violating $R$-Parity at the $B$-Factory}
\author{D. Elazzar Kaplan}
\address{
	Department of Physics 1560, 
	University of Washington,
	Seattle, WA 98195-1560
	}
\date{17 March 1997}
\maketitle
\def\thefootnote{\fnsymbol{footnote}}

\begin{abstract}
Supersymmetry without $R$-parity contains new tree-level contributions
to $B$-decays.  A brief summary is
provided of the current experimental bounds on those tree-level contributions
which are relevant to $B$ physics.  Signals to look for are outlined in
the context of CP violation and rare decays.  For the first time, the
signature of general $R$-parity violating models is described.  $B$-factory 
experiments will provide an opportunity to look for this signal.
\end{abstract}
\pagebreak

The Minimal Supersymmetric Standard Model (MSSM) \cite{MSSM} does
not contain the accidental symmetries of baryon number (B)
and lepton number (L) which grace the Standard Model (SM).
Therefore, an ad-hoc symmetry called $R$-parity is often imposed 
to keep these global symmetries intact.
This symmetry assigns a charge of $(-1)^{3B+L+2S}$ to each particle,
where S is the particle's spin.  Particles of the SM are even 
under this symmetry, while their superpartners are odd.  No compelling
theoretical arguments exist for such a symmetry.  Therefore, it behooves us
to examine both the limits that current experimental data puts on $R$-parity
violating ($\rp$) couplings and the effects they could have on 
the upcoming experiments on $B$-mesons 
(BaBar, BELLE, HERA B, CLEO, RUN II at FNAL) \cite{Bfac}.  While it has 
already been pointed out that a specific $\rp$ 
model could effect CP violation measurements \cite{Grossman}, this article 
is a comprehensive study of the effects of general models and points out new 
signals not yet discussed in the literature.

The gauge-invariant $\rp$ terms that could be added to the MSSM
are
\be
\lambda_{ijk}L_{i}L_{j}E^c_{k}+\lambda'_{ijk}L_{i}Q_{j}D^c_{k}
+\lambda''_{ijk}U^c_{i}D^c_{j}D^c_{k},
\ee
in which bilinear terms are assumed to be rotated away \cite{Hall}.  The 
couplings $\lambda$ and $\lambda''$ are antisymmetric in their first two and
last two (flavor) indices, respectively.  Thus, there are fourty-five
new independent terms and couplings.  Bounds on the proton lifetime
do not allow for reasonably-sized couplings for all of the terms 
\cite{proton,proton2}.  However, if we impose $B$ or $L$ conservation, 
the remaining couplings are much more weakly bounded.
From these bounds, we can calculate the maximum possible effect
on $B$-meson decays and proceed to look for these effects at the 
$B$-factories \cite{Bfac}.

In the last few years, upper limits have been put on many
of the $\rp$ couplings using current experimental data.
Relatively strong bounds have been put on both $\lambda$
and $\lambda'$ couplings, but few similar bounds have been put
on the $\lambda''$ coupling.  This is due mainly to the
difficulty in measuring exclusive hadronic decays.  The upper bounds
on all $\lambda$ couplings range between .04 and .1 for 
slepton masses of $100\GeV$.  They could lead to small amounts of CP violation,
and contribute to rare leptonic decays.  I will not focus on them \cite{lam}.
The majority of the 
$\lambda'$ couplings are most strongly bound by their contribution to the 
decay $K\to\pi\nu\nu$.  The remaining couplings are bounded less
strongly by a number of different experimental phenomena.
In the case of the $\lambda''$ couplings, only two are bounded
significantly below unity.  These are $\lambda''_{112} \leq 10^{-6}$ and
$\lambda''_{113} \leq 10^{-4}$, where the former is due to non-observation
of double-nucleon decay and the latter to non-observation of $n-\bar{n}$
oscillations.  Table 1 in \cite{minirev} includes a complete list of
these bounds and their sources.

More stringent bounds have been put on some products of these couplings
\cite{Moh,Choud,Jang}.  Bounds associated with $K$--$\bar K$ mixing, 
$B$--$\bar B$ mixing and neutrinoless double beta decay limit possible 
effects of $\rp$ terms on $B$-meson decays.  These bounds are as follows:
\begin{eqnarray}
{\rm Re}\left[\sum_{i,j,j'}\left(\frac{100 \GeV}{m_{\tilde \nu_i}}\right)^2
   \lambda'^*_{ij3}\lambda'_{ij'1} V^*_{j1} V_{j'3} \right] 
   &\leq&3\times10^{-8}\;\;\;({\rm from}\:\beta\beta_{0\nu},
					\:{\rm and}\:\Delta m_B)\\
{\rm Re}\left[\sum_{i,j,j'}\left(\frac{100 \GeV}{m_{\tilde \nu_i}}\right)^2
   \lambda'^*_{ij2}\lambda'_{ij'1} V^*_{j1} V_{j'2} \right] 
   &\leq&4.5\times10^{-9}\;\;\;({\rm from}\:\Delta m_K),
\end{eqnarray}
where $V_{jk}$ are Cabibo-Kobayashi-Maskawa (CKM) matrix elements.  The 
bounds are shown in the usual mass basis of the right-handed and up-type 
left-handed quarks.  Since these bounds are not rigorous ones on individual 
products, I shall take only their orders of magnitude and assume no 
`conspiratorial' cancellations in the sum.  Despite these bounds, dramatic
phenomenological effects are not ruled out.  Table 1 provides a list of bounds
on the products of $\lambda'$ couplings that directly affect hadronic
$B$-meson decays.  Individual $\lambda''$ bounds are also shown.

In the next few years, $B$-factory experiments will be on-line, allowing 
for testing of many SM predictions.  One such test is of the predicted 
CP violation associated with the phase in the CKM matrix.  The matrix is 
often parametrized by triangles in the complex plane.
Measurements of angles and lengths of sides of these triangles can, 
with varying degrees of accuracy, be extracted from experimental data.

For example, some angles of the triangle shown in Figure 2 of \cite{PDG} 
can be extracted from the asymmetry of $\bd$ and $\bdb$ decays into the same 
CP eigenstate, $f_{CP}$.  This asymmetry is 

\begin{eqnarray}
a_{f_{CP}}(t)&\equiv&\frac{\Gamma(B_{phys}^{0}(t) \to f_{CP}) -
\Gamma(\bar B_{phys}^{0}(t) \to f_{CP})}
{\Gamma(B_{phys}^{0}(t) \to f_{CP}) +
\Gamma(\bar B_{phys}^{0}(t) \to f_{CP})}\nonumber\\
&&\nonumber\\
&=&\frac{(1-|r_{f_{CP}}|^2)\cos{\Delta Mt} 
- 2{\rm Im}(r_{f_{CP}})\sin{\Delta Mt}}
{1 + |r_{f_{CP}}|^2},
\end{eqnarray}

\noindent where $B_{phys}^{0}(t)\:(\bar B_{phys}^{0}(t))$ is the state at
$t=0$ which is the pure flavor eigenstate, $\bd\:(\bdb)$, and
$\Delta M$ is the difference in masses of the two mass eigenstates.
Moreover, $r_{f_{CP}} \equiv q \bar \CA_{f_{CP}} / p \CA_{f_{CP}}$,
where $\CA_{f_{CP}} (\bar \CA_{f_{CP}})$ is the total decay amplitude of
$\bd\:(\bdb)$ into $f_{CP}$, and $p$ and $q$ are $B$--$\bar B$ mixing
parameters for which $|q/p| \simeq 1$.  Each diagram that contributes to the
total decay amplitude has the form $A e^{i(\psi + \phi)}$, where
$A$ is the magnitude of the amplitude and $\phi$ and $\psi$ are the weak
and strong phases respectively.  The weak phase comes from the weak
couplings in the diagram and the strong phase comes from final-state
rescattering effects.  The strong part of the Hamiltonian is
CP-symmetric; the weak part is not.  Thus, the CP conjugate of the above
decay amplitude is $A e^{i(\psi - \phi)}$.  This analysis and notation
follows that of \cite{PDG}.

When only one diagram completely dominates a given
decay amplitude, $|\CA_{\not R_p}| = |\bar\CA_{\not R_p}|$
and $|r_{f_{CP}}| \simeq 1$.  This simplifies $a_{f_{CP}}$ and allows
for immediate extraction of one of the phases in the CKM matrix.
Certain decays give angles in the triangle.  For example, the CKM 
angle $\beta$ can be extracted from the asymmetry of decays into 
$f_{CP} = J/\psi \pi^0$, which are proportional to

\begin{eqnarray}
  {\rm Im}(r_{f_{CP}})
  &=&{\rm Im}\left(\frac{\vtb^*\vtd\vcb^*\vcd}
    {\vtb\vtd^*\vcb\vcd^*}\right)\nonumber\\
  &=&-\sin{2\beta}.
\end{eqnarray}

\noindent This is an example of CP violation due to interference between
mixing and decay.  If, however, more than one diagram contributes to the same
process, and the diagrams have different strong and weak phases, then (e.g.,
for two contributions)
\be
  |r_{f_{CP}}| \simeq 
  \frac{\sqrt{A^2_1 + A^2_2 + 
	2 A_1 A_2 \cos(\psi_1 - \psi_2 + \phi_2 - \phi_1)}}
  {\sqrt{A^2_1 + A^2_2 + 2 A_1 A_2 \cos(\psi_1 - \psi_2 + \phi_1 - \phi_2)}}
  \neq 1.
\ee

Now the relationship between the weak phase and $a_{f_{CP}}$ is not
so straightforward.  This is a case of direct CP violation.  In such a
case, asymmetries could be detected in decays into non-CP eigenstates
(i.e., in $B^0 \to f$ and $\bar B^0 \to \bar f$).
Asymmetries could also be seen in the corresponding $B^\pm$ decays.

The $\rp$ terms will give new tree-level contributions to 
three-body $b$-decays.  The upper limits on contributions
to the quark subprocess shown in Table 2 come from the current bounds on
$\lambda'$ and $\lambda''$.  The $|\CA_{\not R_p}|/|\CA_{SM}|$ 
in the last two columns is the ratio of the amplitudes of the dominant 
quark diagrams, and not of the full hadronic processes, which include 
strong matrix elements of the type:
\be
  {|{\cal M}_{\not R_p}| \over |{\cal M}_{SM}|} =
  {|\langle\Psi K_S | (\bar b^\alpha_R c^\alpha_L)
  (\bar c^\beta_L s^\beta_R) | \bd\rangle | \over
  |\langle\Psi K_S | (\bar b^\alpha_L \gamma^\mu c^\alpha_L)
  (\bar c^\beta_L \gamma_\mu s^\beta_L) | \bd\rangle |}.
\ee
It is not known how to calculate these ratios accurately.  I shall
assume them to be of $\CO(1)$ and limit bounds on contributions to 
order of magnitude estimates.

Nearly all of the decays listed in Table 2 could have significant
contributions from the new physics, but not simultaneously.  Remarkably,
a signal emerges.  An $\rp$ theory could deviate notably from the SM in 
its predictions of decay rates into heavy mesons or decay rates into solely
light mesons, but not both.  Other theories which could allow for new tree
contributions to $B$-decays (e.g., theories with extra Higgs or diquarks) need
not satisfy this constraint.  These deviations from the SM could be detected 
in a variety of ways:
\begin{itemize}%
  \item The SM predicts that the asymmetry in some decays
	are proportional to the same CKM angle.  A difference
	in the asymmetries would imply an additional contribution
	to at least one of the decays.  For example, the SM predicts
	that the asymmetries in $\bd\to\Psi K_S$ and 
	$\bd\to D^+ D^-$ are both proportional to the sine 
	of the angle $\beta$ \cite{PDG}.  If $\rp$ terms contribute 
	significantly to either process with a different weak phase, the 
	asymmetries would differ, indicating new physics.
	Another example is the decays $\bd\to D^+ D^-$
	and $\bd\to\Psi\pi^0$, which have the same quark
	subprocesses.  In this case, if there is a significant
	$\rp$ contribution with a weak phase different from
	the SM contribution, the asymmetries will be different 
	as long as the ratio of strong matrix element contributions 
	differ from each other, i.e., if

	\be
	\frac{\langle D^+ D^- | (\bar b^\alpha_L \gamma^\mu c^\alpha_L)
        (\bar c^\beta_L \gamma_\mu d^\beta_L) | \bd\rangle}
	{\langle D^+ D^- | (\bar b^\alpha_R c^\alpha_L)
	(\bar c^\beta_L d^\beta_R) | \bd\rangle} \neq 
	\frac{\langle\Psi \pi^0 | (\bar b^\alpha_L \gamma^\mu c^\alpha_L)
        (\bar c^\beta_L \gamma_\mu d^\beta_L) | \bd\rangle}
	{\langle\Psi \pi^0 | (\bar b^\alpha_R c^\alpha_L)
	(\bar c^\beta_L d^\beta_R) | \bd\rangle}.
	\ee

  \item	The SM also predicts that certain decays will measure the CKM
	angles $\alpha, \beta,$ and $\gamma$.  The sum of these
	angles is $180^{\rm o}$.  If the sum of the measurements 
	differs from $180^{\rm o}$, this is a signal of new physics.

  \item The SM predicts very little direct CP violation in most 
	tree-level $B$-decays.  For example, there is a single 
	diagram which dominates the SM contribution to 
	$\bd\to D^+_s D^-$ (and its excited modes),
	and therefore, the direct CP violation in this mode is
	expected to be negligible.  A contribution to
	$\bar{b}\to\bar{c}c\bar{s}$ with phases different 
	from the SM contribution could give a measureable
	asymmetry.  This effect would also be seen in $B^\pm$ decays.

  \item The following quark subprocesses do not appear at tree-level
	in the SM:  $\bar{b}\to\bar{d}d\bar{d}, \bar{d}d\bar{s}, 
	\bar{d}s\bar{s}, \bar{s}s\bar{s}, \bar{d}s\bar{d}$ and 
	$\bar{s}d\bar{s}$.  The first four are dominated by penguin 
	diagrams and the last two by 1-loop box diagrams.  In a 
	supersymmetric theory without $R$-parity, any of these 
	decays could be allowed at tree-level.  Within current bounds,
	$B\to \bar{K}^0 K^0,\:\phi\pi^0,\:\phi K^0,\:K^0 K^0$ and 
	$\bar{K}^0\pi^0$ (and their excited states) could be seen in 
	abundances greater than those predicted by the SM
	signaling the existence of new physics.  The last two modes would only
	have to be seen, as they are predicted to be extremely rare by the SM.
	There is, of course, a complementary set of charged $B$-decays:
	$B^+\to K^+ \bar{K}^0,\:\phi\pi^+,\:\phi K^+,\:K^0 \pi^+,\:
	K^+ K^0$ and $\bar{K}^0 \pi^+$.  Again, the last two are greatly
	suppressed in the SM.

\end{itemize}

In addition to hadronic decays, leptonic and semileptonic $B$-decay rates
could be increased greatly by $\rp$ effects.  The process,
$B\to \ell^+\ell^-$, is highly suppressed in the SM when the leptons 
are the same and forbidden by the SM when they are different \cite{Buch}.
The experimental bounds on this process \cite{CLEO} set some of 
the strongest limits on the relevant $\rp$ parameters \cite{Jang}.
The high statistics at the $B$-factories
will make it possible to probe these limits and look for an enhanced
signal or lepton flavor violation.  
The existance of $\rp$ couplings could result in lepton
nonuniversality in $\bar{b}\to \bar{c}\,\ell^+\nu$ and
$\bar{b}\to \bar{u}\,\ell^+\nu$.  Current limits allow contributions
of 5\% and 100\% of the SM rates respectively.  Contributions to 
one choice of $\ell$ at these levels would lead to
detectable violation of universality 
at the $B$-factories.  Enhancements of $B\to X\bar\nu\nu$ \cite{Gross2} 
could also be large, but may be difficult to detect in experiments 
of the near future \cite{Cola}. See reference \cite{Ali} for a 
full analysis of $B$-decays.

The effects outlined here are highly suppressed or completely forbidden
in supersymmetric theories with $R$-parity, because they do not exist
at tree-level.  Theories which include additional tree-level contributions 
to $b$-quark decays, such as the SM with diquarks or additional Higgs 
scalars, could have similar effects.  Because such theories are not 
restricted to the same signal pattern as $\rp$ theories, it may be possible
to distinguish them experimentally.

This work has been supported in part by the DOE under grant 
\#DE-FG03-96ER40956.  I thank Ann Nelson for numerous useful conversations,
Martin Savage, Steven Wasserbaech and David Wright for helpful discussions,
Nic Nigro for his editing and Laura for her support.

\pagebreak
\begin{tabular}{|c|r|c|c|r|c|}
\hline
\multicolumn{6}{|c|}
  {Bounds on $\sum_{i=1}^{3} |\lambda'^*_{ijk} \lambda'_{ij'k'} |$}\\
\hline
\hfil ($j,k;j',k'$) \hfil&\hfil Upper Bound \hfil&\hfil Source \hfil&
\hfil ($j,k;j',k'$) \hfil&\hfil Upper Bound \hfil&\hfil Source \hfil\\
\hline
(1,3;1,1)&$2\times 10^{-5}(\frac{\mn}{100\GeV})^2$&(a)&
   (3,1;1,1)&$3\times 10^{-3}\footnotemark[1]$&(b),(d)\\
(1,3;1,2)&$1.4\times 10^{-4}\frac{\mb\ms}{(100\GeV)^2}$&(b)&
   (3,1;1,2)&$1\times 10^{-7}(\frac{\mn}{100\GeV})^2$&(c)\\
(1,3;2,1)&$9\times 10^{-7}(\frac{\mn}{100\GeV})^2$&(a)&
   (3,1;2,1)&$3\times 10^{-3}\footnotemark[1]$&(b),(d)\\
(1,3;2,2)&$1.4\times 10^{-4}\frac{\mb\ms}{(100\GeV)^2}$&(b)&
   (3,1;2,2)&$6\times 10^{-7}(\frac{\mn}{100\GeV})^2$&(c)\\
(2,3;1,1)&$8\times 10^{-5}(\frac{\mn}{100\GeV})^2$&(a)&
   (3,2;1,1)&$3\times 10^{-6}(\frac{\mn}{100\GeV})^2$&(c)\\
(2,3;1,2)&$1.4\times 10^{-4}\frac{\mb\ms}{(100\GeV)^2}$&(b)&
   (3,2;1,2)&$5\times 10^{-3}\footnotemark[1]$&(b),(d)\\
(2,3;2,1)&$4\times 10^{-6}(\frac{\mn}{100\GeV})^2$&(a)&
   (3,2;2,1)&$1\times 10^{-6}(\frac{\mn}{100\GeV})^2$&(c)\\
(2,3;2,2)&$1.4\times 10^{-4}\frac{\mb\ms}{(100\GeV)^2}$&(b)&
   (3,2;2,2)&$5\times 10^{-3}\footnotemark[1]$&(b),(d)\\
\hline
\hline
\multicolumn{6}{|c|}
  {Bounds on $|\lambda''_{ijk}|$}\\
\hline
$\lambda''_{112}$&$10^{-6}\footnotemark[2]$&(w)
&$\lambda''_{113}$&$10^{-4}\footnotemark[3]$&(x)\\
$\lambda''_{3jk}$&0.97\footnotemark[1]&(y)
&all others&1.25&(z)\\
\hline
\end{tabular}
\vskip .2in
Table 1.  Product bounds on $\rp$ couplings from the following
   sources:  (a) $\beta\beta_{0\nu}$ \cite{Moh} and $B$--$\bar{B}$ 
   mixing \cite{Choud}, (b) $K\rightarrow\pi\nu\nu$ \cite{Gra}, 
   (c) $K$--$\bar{K}$ mixing \cite{Choud}, (d) a combination of Atomic
   parity violation \cite{atom-nu}, $\nu_\mu$ deep-inelastic 
   scattering \cite{atom-nu}, Z decay width \cite{zdec},
   and top-decay.  The bounds on $\lambda''$ come from:  (w) double
   nucleon decay \cite{doubn}, (x) n--$\bar {\rm n}$ oscillations \cite{doubn},
   (y) Z decay width \cite{zdec2}, (z) perturbative unitarity 
   \cite{proton2,pertu}.
\footnotetext[1]{These terms are linearly dependent on different squark 
		and/or slepton masses.  The values shown are for
		$\tilde m=100\GeV$}
\footnotetext[2]{This bound is proportional to the square of the squark
		masses, keeping the gluino mass fixed.  The value shown
		is for $\tilde m=100\GeV$}
\footnotetext[3]{This bound grows exponentially with squark masses.  The
		value shown is for $\tilde m=100\GeV$.  At $\tilde m=400\GeV$,
		it reduces to $\,\sim\!\! 10^{-2}$.}

\pagebreak
\begin{tabular}{|l|c|c|c|c|c|}
\hline
\hfil Quark \hfil&\hfil $B^0_d$ Decay \hfil&
\hfil $B^0_d$ Decay \hfil&\hfil $B^+_d$ Decay \hfil&
\multicolumn{2}{c|}
	{\hfil Upper Bound on $\frac{|\CA_{\not R_p}|}{|\CA_{SM}|}$\hfil}
\\ \cline{5-6}
\hfil Process \hfil&\hfil Products (CP)\hfil&
\hfil Products (\,$\not\!\!\!{\rm CP}$)\hfil&
\hfil Products \hfil&
\hfil With $\lambda'$ \hfil&\hfil With $\lambda''$ \hfil\\
\hline
$\bar{b} \rightarrow \bar{c}c\bar{s}$&
${\rm J}/\Psi K_S, \Psi (2\rm S) K_S$&
$D^+_s D^-, D^+_s D^{*-}$&$\Psi K^+, D^+_s \bar D^0$&
$10^{-2} (\frac{\mb\ms}{\me^{2}})$&
$10^{2} (\frac{100 \GeV}{\md})^2$
\\ [3pt]
\hline
$\bar{b} \rightarrow \bar{c}c\bar{d}$&
$D^{+}D^{-}, {\rm J}/\Psi \pi^0$&
${\rm J}/\Psi \rho^0, D^{*+} D^-$&$\Psi \pi^+, D^+ \bar D^0$&
$10^{-3} (\frac{\mn}{\me})^2$&
$10^{2} (\frac{100 \GeV}{\ms})^2$
\\ [3pt]
\hline
$\bar{b} \rightarrow \bar{c}u\bar{d}$&
$D_{CP}\pi^0, D_{CP}\rho^0$&
$D^- \pi^+, D^{*-} \pi^+$&$\bar D^0 \pi^+, \bar D^0 \rho^+$&
$10^{-3} (\frac{\mn}{\me})^2$&
$10^{2} (\frac{100 \GeV}{\ms})^2$
\\ [3pt]
$\longrightarrow\bar{u}c\bar{d}$&
"&$D^+ \pi^-, D^{*+} \pi^-$&$D^+ \pi^0, D^0 \pi^+$&
$10^{-2} (\frac{\mn}{\me})^2$&
$10^{-2}$
\\ [3pt]
\hline
$\bar{b} \rightarrow \bar{c}u\bar{s}$&
$D_{CP} K_S$&$D_{CP} K^{*0}$&$\bar D^0 K^+, \bar D^0 K^{*+}$&
$10^{-1} (\frac{\mb\ms}{\me^{2}})$&
$10^{-2}\footnotemark[4]$
\\ [3pt]
$\longrightarrow\bar{u}c\bar{s}$&
"&$D^+_s \pi^-, D^+_s \rho^-$&$D^0 K^+,  D^+_s \pi^0$&
$10^{-1} (\frac{\mb\ms}{\me^{2}})$&
$10^{-3}$
\\ [3pt]
\hline
$\bar{b} \rightarrow \bar{u}u\bar{d}$&
$\pi \pi, \pi^0 \rho^0, ...$&
$\rho\rho,\pi^+\rho^-$&$\pi^+\pi^0, \pi^+ \rho^0,...$&
$10^{-2} (\frac{\mn}{\me})^2$&
$10^{-3}$
\\ [3pt]
$\longrightarrow\bar{d}d\bar{d}$&
"&$\rho^0\rho^0$&"&
10\footnotemark[1]&---
\\ [3pt]
\hline
$\bar{b} \rightarrow \bar{u}u\bar{s}$&
$K_S \pi^0, K_S \rho^0$&
$K^+ \pi^-, K^{*+} \pi^-$&$K^+ \pi^0, K^+ \rho^0$&
$1 (\frac{\mb\ms}{\me^{2}})$&
$10^{-6}$
\\ [3pt]
$\longrightarrow\bar{d}d\bar{s}$&
"&$K^{*0} \pi^0, K^{*0} \rho^0$&$K^+ \pi^0, K^0 \pi^+$&
10\footnotemark[2]&
$10^{-6}\footnotemark[2]$
\\ [3pt]
$\longrightarrow\bar{d}s\bar{d}$&
"&"&$\bar K^0 \pi^+, K^{*0} \pi^+$&
$\gg 1$\footnotemark[3]&---
\\ [3pt]
\hline
$\bar{b} \rightarrow \bar{d}s\bar{s}$&
$\phi \pi^0, K_S K_S$&
$\phi \rho^0, \bar K^0 K^{*0}$&$\phi\pi^+, K^+ \bar K^0$&
$\gg 1$\footnotemark[3]&
$\gg 1$\footnotemark[3]  
\\ [3pt]
$\longrightarrow\bar{s}d\bar{s}$&
$K_S K_S$&$K^0 K^{*0}, K^{*0} K^{*0}$&$K^+ K^{*0}, K^{*+} K^0$&
$\gg 1$\footnotemark[3]&---
\\ [3pt]
\hline
$\bar{b} \rightarrow \bar{s}s\bar{s}$&
$\phi K_S$&
$\phi K^{*0}$&$\phi K^+$&
$\gg 1$\footnotemark[3]&---
\\ [3pt]
\hline
\end{tabular}
\vskip .2 in
Table 2.  Bounds on $R$-parity violating contributions to $B$-decays.
		Sparticle masses are assumed to be $100\GeV$ where not
		shown.\\
\footnotetext[1]{Compared with $\bar{b} \rightarrow \bar{u}u\bar{d}$ 
			in the SM.}
\footnotetext[2]{Compared with penguin.} 
\footnotetext[3]{Bound dominates the 1-loop SM contribution.}
\footnotetext[4]{This bound becomes unity at $m_{sparticles}=400\GeV$.}


\begin{thebibliography}{99}
\bibitem{lam} The $\lambda$ couplings could contribute to CP violation
   in the $B$ system if some $\lambda'$ couplings were non-zero.  However,
   under current bounds, asymmetries in $B$-decays due to these should
   be undetectable in experiments of the near future.  The products of
   coupling, $\lambda\lambda'$ and $\lambda' \lambda'$, could both
   contribute to rare leptonic and semi-leptonic $B$-decays.
   The bounds on $\lambda' \lambda'$ are weaker than those on
   $\lambda\lambda'$ in most relevant cases.
\end{thebibliography}
\end{document}